\documentclass[epj]{webofc}
\usepackage[utf8]{inputenc}
\usepackage[varg]{txfonts}   
\usepackage{booktabs}
\usepackage{xcolor}
\definecolor{darkred}{rgb}{0.4,0.0,0.0}
\definecolor{darkgreen}{rgb}{0.0,0.4,0.0}
\definecolor{darkblue}{rgb}{0.0,0.0,0.4}
\usepackage[bookmarks,linktocpage,colorlinks,
    linkcolor = darkred,
    urlcolor  = darkblue,
    citecolor = darkgreen]{hyperref}
\newcommand{\be}{\begin{equation}}
\newcommand{\ee}{\end{equation}}
\newcommand{\bea}{\begin{eqnarray}}
\newcommand{\eea}{\end{eqnarray}}
\newcommand{\MSbar}{{\overline{\rm MS}}}

\usepackage{subfigure}
\wocname{EPJ Web of Conferences}
\woctitle{Lattice2017}
\begin{document}
%
\selectlanguage{english}
\title{%
Perturbative Renormalization of Wilson line operators 
}
\author{%
\firstname{Martha} \lastname{Constantinou}\inst{1} \and
\firstname{Haralambos} \lastname{Panagopoulos}\inst{2}\fnsep\thanks{Speaker, \email{haris@ucy.ac.cy}}
}
\institute{%
Department of Physics,  Temple University,  Philadelphia,  PA 19122 - 1801,  USA
\and
Department of  Physics,  University  of Cyprus,  POB  20537,  1678  Nicosia,  Cyprus
}
\abstract{%
We present results for the renormalization of gauge invariant nonlocal
fermion operators which contain a Wilson line, to one loop level in
lattice perturbation theory. Our calculations have been performed for
Wilson/clover fermions and a wide class of Symanzik improved gluon
actions. 
The extended nature of such `long-link' operators results in a
nontrivial renormalization, including contributions which diverge
linearly as well as logarithmically with the lattice spacing, along
with additional finite factors. We
present nonperturbative prescriptions to extract the linearly
divergent contributions.  
}
\maketitle
\section{Introduction}\label{intro}

Parton distribution functions (PDFs) provide important information on
the quark and gluon structure of hadrons; at leading twist, they
give the probability of finding a specific parton in the hadron
carrying certain momentum and spin, in the infinite momentum frame.
Due to the fact that PDFs are light-cone correlation functions, they
cannot be computed directly on a Euclidean lattice. Nevertheless, there is an alternative
approach, proposed by X. Ji, involving the computation of quasi-distribution functions, which are
accessible in Lattice QCD. Exploratory studies of the quasi-PDFs reveal 
promising results for the non-singlet operators in the unpolarized, helicity
and transversity cases.

A standard way of extracting quasi-distribution functions in lattice simulations involves computing hadronic matrix elements of certain 
gauge-invariant nonlocal operators; these are made up of a product of an anti-quark field at position $x$, 
possibly some Dirac gamma matrices, a path-ordered exponential of the gauge field (Wilson line) along a path
joining points $x$ and $y$, and a quark field at position $y$. 
Given the extended nature of such operators, an endless variety of them, with 
different quantum numbers, can be defined and studied in lattice simulations and in phenomenological models. In pure gauge theories, 
prototype nonlocal operators are path-ordered exponentials along closed contours (Wilson loops); the contours may be smooth, but they may 
also contain angular points (cusps) and self-intersections. 

The history of investigations of nonlocal operators in gauge theories
goes back a long time.
In particular, the renormalization of Wilson loops
has been studied perturbatively, in dimensional regularization (DR).
Using arguments valid to all orders in perturbation theory, it was shown that smooth Wilson loops
in DR are finite functions of the renormalized coupling, while the presence of cusps and
self-intersections introduces logarithmically divergent multiplicative renormalization factors; at
the same time, it was shown that other regularization schemes are expected to lead to further
renormalization factors $Z$ which are linearly divergent with the dimensionful ultraviolet cutoff
$a$:
\be
Z = e^{\displaystyle - c\,L/a}\,,
\ee
where $c$ is a dimensionless function of the renormalized coupling $g$, and $L$ is the loop length.

There are several obstacles which need to be overcome before a transparent picture of PDFs 
can emerge via Ji's approach; one such obstacle is clearly the intricate renormalization behavior, 
which is the object of our present study.
In what follows we formulate the problem, providing the
definitions for the operators which we set out to renormalize, along with
the renormalization prescription. Our calculations are performed both in dimensional
regularization and on the lattice; we address in detail new features appearing on the
lattice, such as contributions which diverge linearly and logarithmically with the lattice spacing,
and finite mixing effects allowed by hypercubic symmetry. We also provide a 
prescription for estimating the linear divergence using non-perturbative data and extending arguments
from perturbation theory. Finally, we point out some
open questions for future investigations. A long write-up of our work,
together with an extended list of references, can be found
in~\cite{Constantinou:2017sej} (see also the companion paper~\cite{Alexandrou:2017huk}).

\section{Formulation}\label{sec-2}

In our lattice calculations we make use of the clover (Sheikholeslami-Wohlert)
fermion action; we allow the clover parameter, 
$c_{\rm SW}$, to be free throughout. 
We are interested in mass-independent renormalization schemes, and
therefore we set the Lagrangian masses for each flavor, $m^f_0$\,, to their critical value;
for a one-loop calculation this corresponds to $m^f_0{=}0$.
In the gluon sector we employ a 3-parameter family of Symanzik-improved actions involving
Wilson loops with 4 and 6 links; members of this family are the
Wilson, Iwasaki and
tree-level Symanzik-improved actions.

The operators which we study in this work have the general form:
\be
\mathcal{O}_\Gamma\equiv \overline\psi(x)\,\Gamma\,\mathcal{P}\,
e^{i\,g\,\int_{0}^{z}  A^\mu(x+\zeta\hat{\mu}) d\zeta}\, \psi(x+z\hat{\mu})\,,
\label{Oper}
\ee
with a Wilson line of length $z$ inserted between the fermion fields in order to ensure gauge invariance.
The appearance of contact terms beyond
tree level renders the limit $z\to 0$ nonanalytic.
We consider only cases where the Wilson line is a straight line along
the $\mu$-axis. Without loss of generality we choose $\mu = 1$. 
We perform our calculation for all independent combinations of Dirac matrices, $\Gamma$, that is:
\vspace*{.25cm}
\begin{equation}
\Gamma = \hat{1},\quad \gamma^5,\quad \gamma^\nu,\quad \gamma^5\,\gamma^\nu,\quad  \sigma^{\nu\rho}.
\end{equation}
In the above, $\rho\ne\mu$ and we distinguish between the cases in which the index $\nu$ is in the same direction as the Wilson line ($\nu=\mu$),
or perpendicular to the Wilson line ($\nu\neq\mu$). The 16 possible choices of $\Gamma$ are separated
into 8 subgroups, whose renormalization is a priori different, defined
as follows:  
\be
\label{S}
\begin{array}{lcllcllcllclrcl}
  {S} &\equiv& \mathcal{O}_{\hat{1}}          & {V}_1 &\equiv&
  \mathcal{O}_{\gamma^1}  & {V}_\nu &\equiv& \mathcal{O}_{\gamma^\nu}
  & {T}_{1\nu} &\equiv& \mathcal{O}_{\sigma^{1\nu}}  \,\quad\,\,\,
  &(\nu, \rho& = &2, 3, 4)       \\
  {P} &\equiv& \mathcal{O}_{\gamma^5}\ \ \            &
  {A}_1 &\equiv& \mathcal{O}_{\gamma^5\gamma^1}\ \ \    &  {A}_\nu
  &\equiv& \mathcal{O}_{\gamma^5\gamma^\nu}\ \ \  &
  {T}_{\nu\rho} &\equiv& \mathcal{O}_{\sigma^{\nu\rho}} \,.
\end{array}
\ee

We perform the calculation in both dimensional (DR) and lattice
(LR) regularizations, which allows one to extract the lattice renormalization 
functions directly in the continuum $\MSbar$-scheme. 

In the LR calculation we encounter
finite mixing for some pairs of operators (see Subsection~\ref{resultsLR}),  and, thus, here we provide the renormalization prescription in the presence of mixing between two structures, $\Gamma_1$ and $\Gamma_2$,
where one has, a $2\times2$ mixing matrix\footnote{All renormalization functions, generically labeled $Z$, depend
on the regularization $X$ ($X$ = DR, LR, etc.) and on the renormalization scheme $Y$ ($Y$ = $\MSbar$, RI', etc.) and
should thus properly be denoted as: $Z^{X,Y}$, unless this is clear from the context.} ($Z$). 
More precisely, we find mixing within each of the pairs: $\{S,\, V_1\}$, $\{A_2,\, T_{34}\}$, $\{A_3,\, T_{42}\}$, $\{A_4,\, T_{23}\}$, 
in the lattice regularization.
In these cases, the renormalization of the operators is then given by a set of 2 equations:
\begin{equation}
  \binom{{\cal O}_{\Gamma_1}^R}{{\cal O}_{\Gamma_2}^R} =
  \begin{pmatrix} Z_{11} & Z_{12} \\ Z_{21} & Z_{22} \end{pmatrix}^{-1}
  \binom{{\cal O}_{\Gamma_1}}{{\cal O}_{\Gamma_2}}\,.
\end{equation}
Once the mixing matrix $Z_{ij}$ is obtained through the perturbative
calculation of certain amputated Green's functions, as shown
below, it can be applied to non-perturbative bare Green's functions derived from lattice simulation data, in order to
deduce the renormalized, disentangled Green's functions for each of the two operators separately~\cite{Alexandrou:2017huk}.

The one-loop renormalized Green's function of operator
$\mathcal{O}_{\Gamma_i}$ can be obtained from the one-loop bare  
Green's function of $\mathcal{O}_{\Gamma_i}$ and the tree-level Green's function of  $\mathcal{O}_{\Gamma_j}$ ($j\ne
i$); this can be seen starting from the general expression:
\be
\langle \psi^R\,{\cal O}^R_{\Gamma_i}\,\bar \psi^R \rangle =
Z_\psi\,\sum_{j=1}^2 \, (Z^{-1})_{ij} \,\langle\psi\,{\mathcal O}_{\Gamma_j}\,\bar \psi \rangle \,,
\qquad \psi  = Z^{1/2}_\psi \psi^R\,,
\label{eq1}
\ee
where the matrix $Z$ and the fermion field renormalization $Z_\psi$ have the following
expansion: 
\be
Z_{ij} = \delta_{ij} + g^2 z_{ij} + \mathcal{O}(g^4)\,,\qquad 
Z_\psi = 1 + g^2 z_\psi + \mathcal{O}(g^4)\,.
\ee

Once the $\MSbar$ Green's functions have been computed in
DR, the condition for extracting $Z^{LR,\,\MSbar}_{ij}$ is simply the requirement that renormalized Green's functions be
regularization independent:
\be
\langle \psi^R\,{\cal O}^R_{\Gamma_i}\,\bar \psi^R \rangle^{DR, \,\MSbar} = 
\left.\langle \psi^R\,{\cal O}^R_{\Gamma_i}\,\bar \psi^R \rangle^{LR, \,\MSbar}\right|_{a\to 0}\,.
\ee
Substituting the right-hand side of the above relation by the expression in Eq.~(\ref{eq1}), there follows:

 \be
\hspace*{-0.35cm}
\langle \psi^R\,{\cal O}^R_{\Gamma_1}\,\bar \psi^R \rangle^{DR, \,\MSbar}
-\langle \psi\,{\cal O}_{\Gamma_1}\,\bar \psi \rangle^{LR}=
g^2\, \Big(z_\psi^{LR,\,\MSbar}-z_{11}^{LR,\,\MSbar}\Big) \langle \psi \,{\cal O}_{\Gamma_1}\,\bar \psi \rangle^{\rm tree} 
- g^2\, z^{LR,\,\MSbar}_{12}  \langle\psi\,{\cal O}_{\Gamma_2}\,\bar \psi \rangle^{\rm tree} + \mathcal{O}(g^4).
\label{eq2}
\ee
The Green's functions on the left-hand side of Eq.~(\ref{eq2}) are the main results of this work, 
where $\langle \psi\,{\cal O}_{\Gamma_1}\,\bar \psi \rangle^{DR, \,\MSbar}$ is the renormalized Green's function for ${\cal O}_{\Gamma_1}$
which has been computed in DR and renormalized using the $\MSbar$-scheme, while
$\langle \psi\,{\cal O}_{\Gamma_1}\,\bar \psi \rangle^{LR}$ is the bare Green's function of ${\cal O}_{\Gamma_1}$ in LR. 
The difference between these Green's functions is polynomial in the external momentum (of degree 0, in our case,
since no lower-dimensional operators mix); in fact, verification of this property constitutes a highly nontrivial check
of our calculations.
Thus, Eq.~(\ref{eq2}) is an appropriate definition of the momentum-independent renormalization functions, $Z_{11}$ and $Z_{12}$.

Non-perturbative evaluations of the renormalization functions cannot be obtained directly in the $\MSbar$ scheme;
rather, one may calculate them in some appropriately defined variant of the RI$'$ scheme, and then introduce the corresponding conversion  
factors between RI$'$ and $\MSbar$. Here we propose a convenient RI$'$ scheme which can be 
applied non-perturbatively, similar to the case of the ultra-local fermion composite operators, with due attention to
mixing. Defining, for brevity: $\Lambda_{\Gamma_i} = \langle\psi\,{\mathcal O}_{\Gamma_i}\,\bar \psi \rangle$\,, and denoting the corresponding renormalized Green's functions by $\Lambda_{\Gamma_i}^{{\rm RI}'}$\,,
we require:
\be 
{\rm Tr}\Big[\Lambda_{\Gamma_i}^{{\rm RI}'} \, (\Lambda_{\Gamma_j}^{\rm tree})^\dagger\Big]_{q_\nu = {\bar q}_\nu} =  
{\rm Tr}\Big[\Lambda_{\Gamma_i}^{\rm tree} \, (\Lambda_{\Gamma_j}^{\rm tree})^\dagger\Big] \qquad \left(\ = 12
\delta_{ij}\ \right).
\label{riprime}
\ee
The momentum of the
external fermion fields is denoted by $q_\nu$, and the four-vector
${\bar q}_\nu$ denotes the RI$'$ renormalization scale. We note that the magnitude of $\bar q$ alone is not
sufficient to specify completely the renormalization prescription: Different directions in $\bar q$ amount to different
renormalization schemes, which are related among themselves via finite renormalization factors. In what follows we will
select the RI' renormalization scale 4-vector to point along the direction $\mu=1$ of the Wilson line: $(\bar q, 0, 0, 0)$.

Using Eq.~(\ref{eq1}) we express Eq.~(\ref{riprime}) in terms of bare Green's functions, obtaining:
\be
\frac{1}{12}\, Z_\psi^{LR,{\rm RI}'}\,\sum_{k=1}^2 {(Z^{LR,{\rm RI}'})^{-1}}_{ik} \,  
{\rm Tr}\Big[\Lambda_{\Gamma_k} \, (\Lambda_{\Gamma_j}^{\rm tree})^\dagger\Big] = \delta_{ij}\,.
\label{eq6}
\ee

Eq.~(\ref{eq6}) amounts to four conditions for the four elements of the
matrix $Z^{LR,{\rm RI'}}$. As it was intended, it lends itself to a non-perturbative evaluation of $Z^{LR,{\rm RI'}}$,
using simulation data for $\Lambda_{\Gamma_k}$\,. 

Converting the non-perturbative, RI'-renormalized Green's functions $\Lambda_{\Gamma_i}^{{\rm RI}'}$ to the $\MSbar$
scheme relies necessarily on perturbation theory, given that the very definition of $\MSbar$ is perturbative. We write:
\be
  \binom{{\cal O}_{\Gamma_1}^\MSbar}{{\cal O}_{\Gamma_2}^\MSbar} =  (Z^{LR,\MSbar})^{-1} \cdot (Z^{LR,{\rm RI}'})\cdot
  \binom{{\cal O}_{\Gamma_1}^{{\rm RI}'}}{{\cal O}_{\Gamma_2}^{{\rm RI}'}} \equiv
 (\mathcal{C}^{\MSbar,{\rm RI}'})\cdot
  \binom{{\cal O}_{\Gamma_1}^{{\rm RI}'}}{{\cal O}_{\Gamma_2}^{{\rm RI}'}}.
\label{eq7}
\ee
The conversion factor $\mathcal{C}^{\MSbar,{\rm RI}'}$ is a $2{\times}2$ matrix in this case; it is constant
($q$-independent) and stays finite as the regulator is sent to its limit ($a\to 0$ for LR, $D\to 4$ for DR).
Most importantly, its value is independent of the regularization;
thus, the evaluation of $\mathcal{C}^{\MSbar,{\rm RI}'}$ can be performed in DR, where
evaluation beyond one loop is far easier than in LR; this, in a nutshell, is the advantage of using the RI' scheme as an
intermediary. A further simplification originates from the fact that the DR mixing matrices $Z^{DR,\MSbar}$ and
$Z^{DR,{\rm RI}'}$ are both diagonal (the latter is true because of the specific direction we have chosen for the
renormalization scale 4-vector); as a result, $\mathcal{C}^{\MSbar,{\rm RI}'}$ turns out to be diagonal as well.  
We stress that, unlike the case of ultra-local operators, the conversion factor may (and, in
general, will) depend on the length of the Wilson line and on the
individual components of the RI$'$ renormalization scale four-vector, through the
dimensionless quantities $z{\bar q}_\nu$\,. 

\section{Calculation - Results}\label{sec-3}
\begin{figure}[thb]
\centering
\includegraphics[width=8cm]{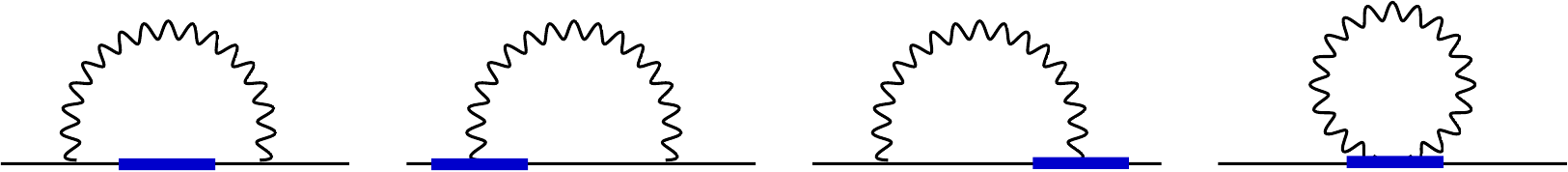}
\caption{Feynman diagrams contributing to the one-loop calculation of the Green's functions of operator $\mathcal{O}_\Gamma$.
The straight (wavy) lines represent fermions (gluons). The operator insertion is denoted by a filled rectangle.}
\label{fig1}
\end{figure}

The Feynman diagrams that enter our one-loop calculations are shown in Fig.~\ref{fig1}, where 
the filled rectangle represents the insertion of a nonlocal operator $\mathcal{O}_\Gamma$\,.
These diagrams will appear in both
LR and DR, since all vertices are present in both regularizations, and since even the ``tadpole'' diagram does not
vanish in DR, by virtue of the nonlocal nature of $\mathcal{O}_\Gamma$\,.
However, the LR calculation is much more challenging: The vertices of $\mathcal{O}_\Gamma$ are more complicated, and
extracting the singular parts of the Green's functions is a more
lengthy and subtle procedure.

\subsection{Dimensional Regularization}
\label{resultsDR}
The perturbative calculation in $D= 4-2\epsilon$ dimensions has been performed in an arbitrary covariant gauge in order to see first-hand 
the gauge invariance of the renormalization functions, as a consistency check.
Pole parts in $\epsilon$ are multiples of the tree-level values, which indicates
no mixing between operators of equal or lower dimension in the $\MSbar$ scheme. Another important characteristic of the 
${\cal O}(g^2)/\epsilon$ contributions is that they are operator independent, in terms of both the Dirac structure 
and the length of the Wilson line, $z$. 
We find a gauge independent renormalization function for the operators
of Eq.~(\ref{Oper}), in agreement with old results:
\begin{equation}
Z^{DR,\,\MSbar}_\Gamma = 1 + \frac{3}{\epsilon}\, \frac{g^2\, C_f}{16\,\pi^2}\,,
\end{equation}

While the independence of $Z^{DR,\,\MSbar}_\Gamma$ from the Dirac matrix insertion $\Gamma$ is a feature valid
to one-loop level, its independence from the length of the Wilson line $z$ is expected to hold to all orders in
perturbation theory; this, in essence, is due to the fact that the most dominant pole at every loop can depend neither on the
external momenta nor on the renormalization scale, thus there is no dimensionless $z$-dependent factor that could appear
in the pole part.

The Green's functions in DR are essential for the computation of the conversion factors between 
different renormalization schemes, and here we are interested in the RI$'$ scheme defined in Eq.~(\ref{eq6}). 
The conversion factor is
the same for each of the following pairs: ($S,\ P$),
($V_1,\ A_1$), ($V_\nu,\ A_\nu$), ($T_{1\nu},\ T_{\nu\rho}$).
The general expressions for $\mathcal{C}_\Gamma $ involve integrals over modified Bessel functions; they are written out explicitly in~\cite{Constantinou:2017sej}. The conversion factors depend on the dimensionless quantities $z\bar q$ and
$\bar q/\bar\mu$, where the RI' and $\MSbar$ renormalization scales ($\bar q$ and $\bar\mu$, respectively) have been left independent.

The Green's functions of operators with a Wilson line are
complex-valued; this property holds also for the non-perturbative matrix
elements between nucleon states and for the conversion factors.

\begin{figure}[thb]
\centering
\includegraphics[width=5cm]{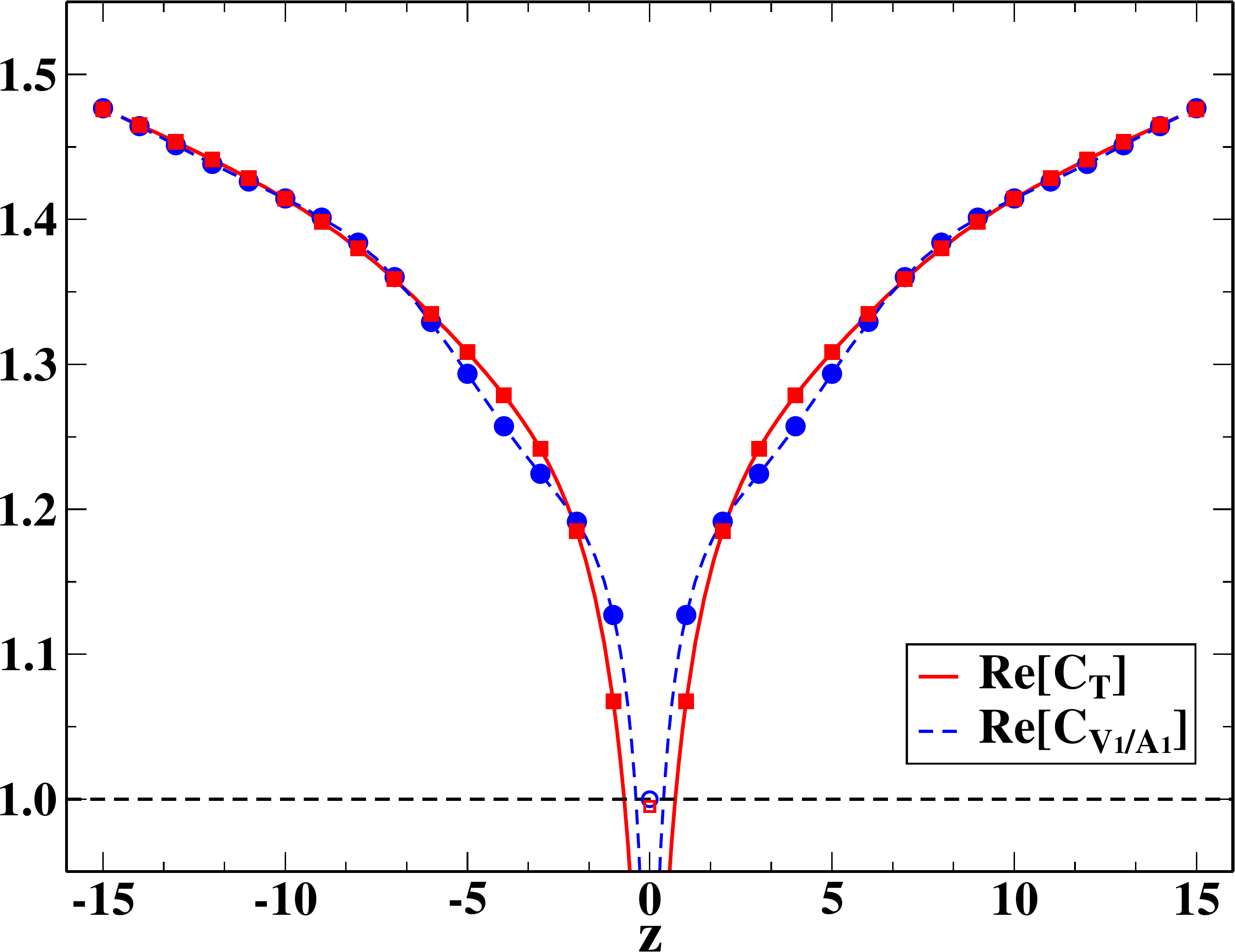}\quad
\includegraphics[width=5cm]{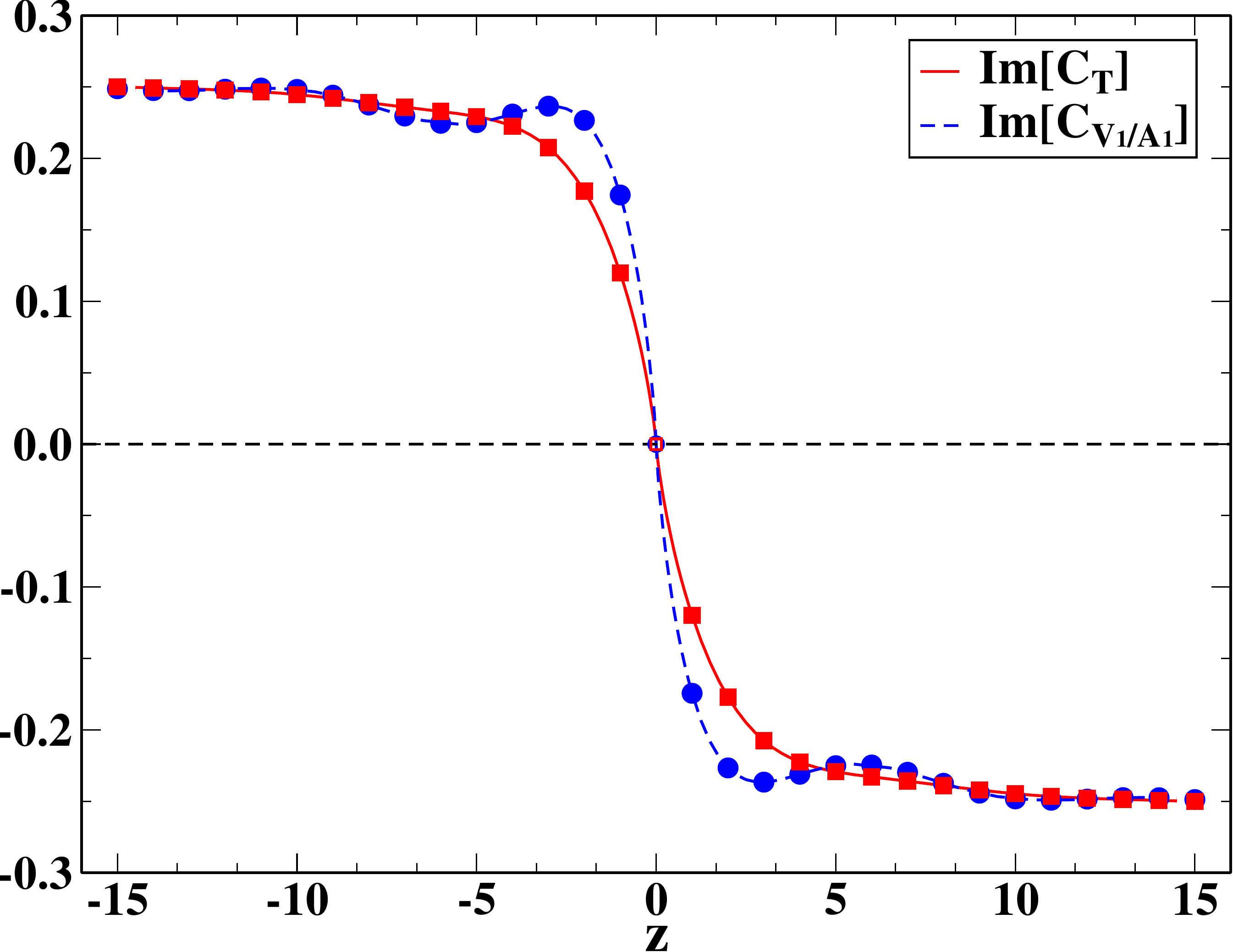}
\caption{\small{Real (left panel) and imaginary (right panel) parts of the conversion factors for the operators $V_1$ and $T$ as a function of $z/a$ in the Landau gauge. 
The RI' momentum scale employed is $a\bar q=\frac{2\pi}{32}\left(4{+}\frac{1}{4},0,0,4\right)$.}}
\label{fig2}
\end{figure}

Note that for a scale of the form $(\bar q,0,0,0)$ the one-loop Green's functions are a multiple of the 
tree-level value of the operator under consideration. 
It is interesting to plot the conversion factors for the cases used in simulations, that is,
$C_{V_1(A_1)}$ and $C_T$. For convenience we choose the coupling constant and the 
RI' momentum scale to match the ensemble of twisted mass fermions employed in 
Ref.~\cite{Alexandrou:2016jqi}:
$g^2{=}3.077$, $a{=}0.082$fm, lattice size: $32^3\times64$ and $a\bar q{=}\frac{2\pi}{32} (\frac{n_t}{2}{+}\frac{1}{4},0,0,n_z)$, 
for $n_t{=}8$ and $n_z{=}4$ (the nucleon is boosted in the $z$ direction). The $\MSbar$ scale is set to $\bar\mu = 2{\rm GeV}$. The conversion 
factors are gauge dependent and we choose the Landau gauge which is mostly used in non-perturbative renormalization. 
In Fig.~\ref{fig2} we plot the real (left panel) and imaginary (right panel) parts of $C_{V_1(A_1)}$ and $C_T$,
as a function of $z/a$. 
One observes that for large values of $z$ the dependence of the conversion factor on the choice of operator becomes
milder. However, this behavior is not granted at higher loops.

\subsection{Lattice Regularization}
\label{resultsLR}

We now turn to the evaluation of the lattice-regularized bare Green's functions $\langle \psi\,{\cal O}_{\Gamma_1}\,\bar
\psi \rangle^{LR}$\,; this is
a far more complicated calculation, as compared to dimensional regularization, because the extraction of the divergences
is more delicate. 
Despite the complexity of the bare Green's functions, their 
difference in DR and LR is necessarily independent of the external momentum $q$, which leads to a
prescription for extracting $Z_\mathcal{O}^{LR,\MSbar} $ (potentially $z$-dependent!) without an intermediate (e.g., RI$'$-type) scheme. 

By analogy with closed Wilson loops in regularizations other than DR, we find a linear
divergence also for Wilson line operators in LR; it is proportional to $|z|/a$ and arises from the 
tadpole diagram, with a proportionality coefficient which depends solely on the choice of the gluon action.

Just as with other contributions to the bare Green's function, the linear divergence is the same -- at one-loop level -- for
all operator insertions. In a resummation of all orders in perturbation theory, the powers of $|z|/a$ are expected to
exponentiate as: $\Lambda_{\Gamma} = \exp(-c \, |z|/a) \, {\tilde\Lambda}_\Gamma \,$,
where $c$ is a power series in coupling, and ${\tilde\Lambda}_\Gamma$ is related to
$\Lambda_\Gamma^\MSbar$ by a further renormalization factor which is at most logarithmically divergent with $a$.
To one loop, we find the following form for the difference between the bare lattice Green's functions and the
$\MSbar$-renormalized ones ($\beta = 1-\alpha$ is the gauge parameter): 
\bea
\langle \psi\,{\cal O}_{\Gamma}\,\bar \psi \rangle^{DR, \,\MSbar}
-\langle \psi\,{\cal O}_{\Gamma}\,\bar \psi \rangle^{LR} = 
\frac{g^2\,C_f}{16\,\pi^2}\, e^{i\,q_\mu z}\, \hspace*{-0.4cm}&\Bigg[&\hspace*{-0.3cm}\Gamma \Big(\alpha_1 + \alpha_2\, \beta + \alpha_3\,\frac{|z|}{a}  
+ \log \left(a^2 \bar\mu^2\right)\left(4-\beta\right) \Big)\nonumber\\
&&\hspace*{-0.3cm} +\left(\Gamma\cdot\gamma_\mu + \gamma_\mu\cdot\Gamma \right)\,\Big(\alpha_4 + \alpha_5\,c_{\rm SW}\Big) \Bigg]\,.
\label{diffDRLR}
\eea
Using Eq.~(\ref{diffDRLR}) together with  Eq.~(\ref{eq2}) one can extract the gauge invariant multiplicative renormalization and mixing coefficients
in the $\MSbar$-scheme and LR.
In our calculation $\alpha_3$ is negative, as expected.

A few interesting properties of Eq.~(\ref{diffDRLR}) can be pointed out: 
The contribution $\left(\Gamma\cdot\gamma_\mu {+} \gamma_\mu\cdot\Gamma \right)$ indicates mixing between operators
of equal dimension, which is finite and appears in the lattice regularization. Moreover, this combination vanishes for certain 
choices of the Dirac structure $\Gamma$ in the operator. 
For the operators $P$, $V_\nu$ ($\nu\ne\mu$), $A_\mu$, $T_{\mu\nu}$ ($\nu\ne\mu$) the combination
$\left(\Gamma\cdot\gamma_\mu + \gamma_\mu\cdot\Gamma \right)$  
is zero and only a multiplicative renormalization is required. This has significant impact in the non-perturbative calculation of the
unpolarized quasi-PDFs, as there is a mixing with a twist-3 scalar operator.
Such a mixing must be eliminated using a proper renormalization prescription, ideally non-perturbatively~\cite{Alexandrou:2017huk}.

To one-loop level, the diagonal elements of the mixing matrix (multiplicative renormalization) are the same for all
operators under study, and through Eq.~(\ref{eq2}) one obtains: 
\be
Z_\Gamma^{LR,\MSbar}  =  1 +  \frac{g^2\,C_f}{16\,\pi^2}\, \left( e_1 + e_2 \,\frac{|z|}{a}  
+e_3\,c_{\rm SW} + e_4\,c_{\rm SW}^2
-3 \log \left(a^2  \bar\mu^2\right) \right),
\label{Zmult}
\ee
where the coefficients $e_1 - e_4$ are given in Table~\ref{tab2}, for
different gluon actions. 
As expected, $Z_\Gamma^{LR,\MSbar}$ is gauge independent, and the cancelation of the gauge dependence was numerically 
confirmed up to ${\cal O}(10^{-5})$. This gives an estimate on the accuracy of the numerical loop integrations.
Similar to $Z_\Gamma^{LR,\MSbar}$, the nonvanishing mixing coefficients are operator independent and have the general form:
\be
Z_{12}^{LR,\MSbar}  = Z_{21}^{LR,\MSbar}  =   0 +  \frac{g^2\,C_f}{16\,\pi^2}\, \left( e_5+ e_6 \,c_{\rm SW} \right),
\label{Zmix}
\ee
where $Z_{ij}^{LR,\MSbar}$ ($i\ne j$) is nonzero only for the operator pairs: $\{S,\, V_1\}$, $\{A_2,\, T_{34}\}$, $\{A_3,\,
T_{42}\}$, $\{A_4,\, T_{23}\}$.
The coefficients $e_5$ and $e_6$ 
are shown in Table~\ref{tab2}. 
Given that the strength of mixing depends on the value of $c_{\rm
  SW}$, one may tune the clover parameter in order to eliminate mixing
at one loop. 

\begin{table}[thb]
\small
\centering
\caption{Numerical values of the coefficients $e_1$ - $e_4$ of the multiplicative renormalization functions and $e_5$ - $e_6$ of the 
mixing coefficients for Wilson, tree-level (TL) Symanzik and Iwasaki gluon actions.}
\begin{tabular}{lcccccc}\toprule
 Action  &  $e_1$ &  $e_2$  &  $e_3$  & $e_4$  & $e_5$ &  $e_6$ \\ \midrule
Wilson				&24.3063	&-19.9548	&-2.24887	&-1.39727	&14.4499	&-8.28467		\\
TL Symanzik	\,\,\,	&19.8442	&-17.2937	&-2.01543	&-1.24220	&12.7558	&-7.67356 		\\
Iwasaki				&12.5576	&-12.9781	&-1.60101	&-0.97321	&9.93653	&-6.52764		\\ \bottomrule
\end{tabular}
\label{tab2}
\end{table}

The linear divergence in the lattice-regularized Wilson line operator
requires a careful removal before the continuum limit can be reached in the non-perturbative matrix elements. One way to
eliminate this divergence is to use the estimate of the  
one-loop coefficient $e_2$ of Eq.~(\ref{Zmult}) and subtract it from the non-perturbative matrix elements. 
However, this subtraction only partially removes the divergence, as higher orders still remain and they will dominate in the
$a\to 0$ limit. Thus,
it is preferable to develop a non-perturbative method to extract the linear divergence. 

Our proposal is based on using bare matrix elements of the Wilson line operators from numerical simulations,
denoted by $q(P_3,z)$ in Ref.~\cite{Alexandrou:2016jqi}. We focus on
the helicity (axial) and transversity (tensor); these exhibit no mixing. In non-perturbative calculations, the nucleon is boosted by
momentum $P_3$ along the direction of the Wilson line. Based on the arguments
presented in the previous Subsection we expect that the renormalized matrix elements can depend on $z$
only through the dimensionless quantity 
$P_3 z$. Furthermore, the dependence on the scale $\bar\mu$ in the renormalized matrix element is well defined and involves the anomalous 
dimension ($\gamma_\Gamma$) of the operator: $q^R(P_3 z, P_3/\bar\mu)\propto {\bar\mu}^{-2\gamma_\Gamma}$, which is
matched by the $\bar\mu$ dependence in the renormalization function.  Thus, 
\be
q^R(P_3 z, P_3/\bar\mu) = (P_3/\bar\mu)^{2\gamma_\Gamma} \cdot {\tilde q}^R(P_3 z).
\ee 
Similarly, the function
$Z_\Gamma^{LR,\MSbar}(a\bar\mu, z/a)$, given its expected $\bar\mu$-dependence, will factorize as: 
\be
Z_\Gamma^{LR,\MSbar}(a\bar\mu, z/a) = {\tilde Z}_\Gamma(a\bar\mu)\cdot \hat Z(z/a).
\ee 
The factor $\hat Z(z/a)$ originates exclusively from tadpole diagrams; the one-loop contribution proportional to $e_2$
in Eq.~(\ref{Zmult}) will exponentiate, upon considering higher powers
of $g$, leading to: 
\be \hat Z(z/a) = e^{-\delta m\,|z|/a}, \qquad \delta m = - \frac{g^2\,C_f}{16\,\pi^2}\, e_2 + {\cal O}(g^4). \ee
This is entirely consistent with the exponential behavior $\exp(-\delta m\,|z|/a)$ proven
 for Wilson loops.

We can thus write the ratio of the bare matrix elements for different values of $P_3$ and $z$ as:
\be
\displaystyle
\frac{q(P_3, z)}{q(P'_3, z')} = 
\frac{Z_\Gamma^{LR,\MSbar}(a\bar\mu, z/a) \cdot q^R(P_3 z, P_3/\bar\mu) }
{Z_\Gamma^{LR,\MSbar}(a\bar\mu, z'/a) \cdot q^R(P'_3 z', P'_3/\bar\mu) } =
\frac{e^{-\delta m\,|z|/a} \, \tilde Z_\Gamma(a\bar\mu) \, \left({\displaystyle \frac{P_3}{\bar\mu}}\right)^{2\gamma_\Gamma} \, \tilde q^R(P_3 z) }
{e^{-\delta m\,|z'|/a} \, \tilde Z_\Gamma(a\bar\mu) \, \left({\displaystyle \frac{P'_3}{\bar\mu}}\right)^{2\gamma_\Gamma} \,
  \tilde q^R(P'_3 z')  } \,,
\label{ratio}
\ee
where the one-loop anomalous dimension is $\gamma_\Gamma =  -3 g^2 C_f/(16\pi^2)$ for all operator insertions. 
Choosing $P_3,\,P'_3,\,z,\,z'$ such that $P_3 z = P'_3 z'$, simplifies the ratio considerably:
\be
q(P_3, z)/q(P'_3, z') = 
e^{-\delta m\,(|z|-|z'|)/a} \, (P_3/P'_3)^{-6 g^2 C_f / (16\pi^2) }\,.
\label{RR}
\ee
Thus, by forming this ratio from non-perturbative data, and
by choosing several combinations of $P_3 z = P'_3 z'$, one can fit to extract the coefficient of the linear 
divergence, $\delta m$.
This ratio can be investigated for the helicity and transversity
separately. Since the right-hand side of Eq.~(\ref{RR}) is independent  
of the operator insertion, one expects the same value for the exponential coefficient, up to lattice artifacts. 
We have tested this method with the data of ETMC presented in Ref.~\cite{Alexandrou:2016jqi}, with encouraging results: $\bullet$ The ratio $q(P_3, z)/q(P'_3, z') $ was found to be real for the helicity and transversity
 as expected from Eq.~(\ref{RR}), despite the fact that the matrix elements themselves are complex $\bullet$ 
The analogous ratio for the unpolarized operator, which mixes with the
scalar, leads to a nonzero imaginary part $\bullet$ The extracted
value for the coefficient $\delta m/a$, using different combinations
of $P_3 z$, is consistent within statistical accuracy $\bullet$ Both helicity and transversity give very similar estimates for $\delta m/a$.
 
 \section{Future Work}
 
A natural continuation of this project is the addition of smearing to the fermionic part of the action and/or to the
gauge links of the Wilson line operator. This is important, as modern simulations employ such smearing techniques 
(e.g., stout and HYP) that suppress the power divergence and bring the renormalization functions closer to their
tree-level values. Smearing the operator under study alters its renormalization functions, thus, the same smearing 
must be employed in the renormalization process. 

An extension of this calculation that we intend to pursue, is the evaluation of lattice artifacts to one loop and to all orders
in the lattice spacing, ${\cal O}(a^\infty,\,g^2)$. This has been 
successfully applied to local and one-derivative fermion operators suppressing lattice artifacts from non-perturbative
estimates. For operators with a long Wilson line 
($z {>>} a$) the lattice artifacts are likely to be more prominent, and therefore, such a calculation will be
extremely beneficial for non-perturbative renormalization.

A possible addition to the present work is the two-loop calculation in DR, from which
one can extract the conversion factor between different renormalization schemes, as well as the anomalous 
dimension of the operators.  The conversion factor up to two loops may be applied to non-perturbative data on 
the renormalization functions, to bring them to the $\MSbar$-scheme at
a better accuracy. Nonzero renormalized masses in the conversion
factors, and differences between flavor-singlet and -nonsinglet
operators would further improve accuracy.

Finally, the techniques developed in this work for the renormalization of quasi-PDFs may be inspiring for the
renormalization of Wilson-line fermion operators of different structure, such as staples. 
This will be of high importance for matrix elements of the transverse momentum-dependent parton distributions  
(TMDs) that are currently under investigation for the nucleon and pion in lattice QCD. 

\bigskip\noindent
{\bf{Acknowledgments}}: 
We would like to thank the members of ETMC, in particular Krzysztof Cichy, for fruitful discussions.
MC acknowledges financial support by the U.S. Department of Energy, Office of Science, Office of Nuclear Physics, 
within the framework of the TMD Topical Collaboration, as well as by the National Science Foundation
under Grant No. PHY- 1714407.

\bibliography{lattice2017}

\end{document}